\documentclass[prb,twocolumn,showpacs,preprintnumbers,amsmath,amssymb,floats,longbibliography]{revtex4-2}
\pdfoutput=1
\usepackage{graphicx}
\usepackage{amsmath}
\usepackage{bbold}
\usepackage{epstopdf}
\usepackage{amsbsy}
\usepackage{color}
\usepackage{hyperref}
\usepackage[utf8]{inputenc}
\usepackage{dcolumn}% Align table columns on decimal point
\usepackage{bm}% bold math

\newcommand{\rr}{{\mathbf{r}}}
\newcommand{\R}{{\mathbf{R}}}
\newcommand{\K}{{\mathbf{k}}}

\newcommand{\NNO}{{Nd$_{1-x}$Sr$_x$NiO$_2$}}
\newcommand{\RNO}{{RNiO$_2$}}

\newcommand{\dx}{{$d_{x^2-y^2}$}}
\newcommand{\dz}{{$d_{3z^2 - r^2}$}}

\begin{document}
\title{Electronic Theory for Scanning Tunneling Microscopy Spectra in Infinite-Layer Nickelate Superconductors}
\author{Peayush Choubey$^{1, 2}$ and Ilya M. Eremin$^{1}$}
\affiliation{
$^1$Institut f\"ur Theoretische Physik III, Ruhr-Universit\"at Bochum, D-44801 Bochum, Germany.\\
$^2$Department of Physics, Indian Institute of Technology (Indian School of Mines), Dhanbad, Jharkhand-826004, India.
}

\date{\today}

\begin{abstract}
Recent scanning tunneling microscopy (STM) observation of U-shaped and V-shaped spectra (and their mixture) in superconducting {\NNO} thin films has been interpreted as presence of two distinct gap symmetries in this nickelate superconductor [Gu et al., Nat. Comm. \textbf{11}, 6027 (2020)]. Here, using a two-band model of nickelates capturing dominant contributions from Ni-$3d_{x^2-y^2}$ and rare-earth (R)-$5d_{3z^2 - r^2}$ orbitals, we show that the experimental observation can be simply explained within a pairing scenario characterized by a conventional $d_{x^2-y^2}$-wave gap structure with lowest harmonic on the Ni-band and a $d_{x^2-y^2}$-wave gap with higher-harmonics on the R-band. We perform realistic simulations of STM spectra employing first-principles Wannier functions to properly account for the tunneling processes and obtain V, U, and mixed spectral line-shapes depending on the position of the STM tip within the unit cell. The V- and U-shaped spectra are contributed from Ni and R-bands, respectively, and Wannier functions, in essence, provide position-dependent weighing factors, determining the spectral line-shape at a given intra-unit cell position. We propose a phase-sensitive experiment to distinguish between the proposed $d$-wave gap structure and time-reversal symmetry breaking $d+is$ gap which yields very similar intra-unit cell spectra.   
\end{abstract}

%\pacs{74.20.-z, 74.70.Xa, 74.62.En, 74.81.-g}

\maketitle
\section{Introduction}
\label{sec:Intro}
The recent discovery of superconductivity in thin films of Sr-doped infinite layer nickelates {\RNO} (R = Nd, Pr, La) \cite{Hwang2019, Hwang2020A, Ariando2021,Hwang2021,Nomura2021} has garnered significant attention from condensed matter community. Infinite layer nickelates were originally envisaged as analogs of cuprate high-$T_{c}$ superconductors, owing to the $3d^{9}$ character of Ni$^{+}$ \cite{Anisimov1999}, same as Cu$^{2+}$ in cuprates. However, subsequent first-principles calculations \cite{Pickett2004} showed that, unlike cuprates, nickeltates are multiband systems with significant $c$-axis electronic dispersion and moderate Ni-O hybridization. Recent experiments have further highlighted the difference between these two systems. Parent compounds of infinite layer nickelates do not show any long-range magnetic order \cite{Ikeda2016, Hayward2003, Hayward1999}, however, a recent NMR study suggests antiferrmomagnetic order in Nd$_{0.85}$Sr$_{0.15}$NiO$_2$ powdered samples \cite{Wen2021A}. Further, the parent compounds of superconducting nickelates are bad metals showing a resistivity upturn at low temperatures ($\sim70$K) \cite{Hwang2019, Hwang2020A}, which has been attributed to Kondo-like physics \cite{Zhang2020A}. Furthermore, Hall effect measurements \cite{Hwang2019, Hwang2020A, Hwang2020B, Hwang2020C} and X-ray spectroscopy \cite{Hepting2020} indicate multi-band character of the parent compound. 

The superconducting dome in {\RNO} thin films extends in hole-doping range $\sim(0.12-0.25)$ \cite{Hwang2020B, Hwang2020C} with maximum transition temperature $T_{c}\sim14$K occurring in Pr compound \cite{Hwang2020C}. Although there is no substantive information about the nature of superconducting state yet, a recent STM study \cite{Wen2020B} on superconducting {\NNO} thin films has revealed the presence of (i) V-shaped spectra that fits well to a simple $d$-wave gap function with maximum gap of 3.9meV, (ii) U-shaped spectra resembling an $s$-wave gap with amplitude 2.35meV, and (iii) spectra showing mixture of both. Due to substantial surface roughness of the sample, it was not possible to identify the precise location of the STM tip with respect to the surface atoms. 

On theory front, nickelates have been proposed to harbour unconventional superconductivity as electron-phonon coupling turns out to be too small to produce the observed high $T_c$ \cite{Arita2019}. Weak-coupling theories of spin-fluctuation mediated pairing \cite{Arita2019, Kuroki2020, Thomale2020A, Das2020, Werner2020A} as well as strong-coupling theories based on $t-J$ like models \cite{Wu2019, Vishwanath2020, Held2020, Zhang2020B} employing multi-band model of the normal state, as predicted by a number of first-principles studies (plane DFT \cite{Pickett2004, Hepting2020, Kuroki2020, Botana2020A, Weng2020, Zhong2019, Devereaux2020, Cano2020A} as well as calculations taking many-body effects into account \cite{Lechermann2020A, Lechermann2020B, Kotliar2020A, Cano2020B, Werner2020B, Han2020, Millis2020, Savrasov2020}), predict $d_{x^2-y^2}$-wave pairing in a large region of parameter space. Note that a cuprate-like simple $d_{x^2-y^2}$-wave pairing will result in a V-shaped spectrum and, hence, can not explain the aforementioned STM results \cite{Wen2020B} on its own.

To the best of our knowledge, there are three existing explanations of the STM results. First, using $K-t-J$ model, Ref. \onlinecite{Zhang2020B} proposed that as a function of doping and Kondo coupling $K$, the superconducing gap evolves from $d$-wave to $s$-wave with intermediate $d+is$-state, and thus, one may obtain  $V$-shaped, $U$-shaped and mixed spectra if the doping and $K$ change substantially over the sample (which is unlikely, as argued in Ref. \cite{Wen2020B}). In this scenario, it is hard to explain (without invoking extreme fine-tuning) why the mixed gap shows kinks and coherence peaks almost at the same energies ($\pm2.35$meV and $\pm3.9$meV, respectively)\cite{Wen2020B} where V-shaped and U-shaped spectra show coherence peaks. Second, Ref. \cite{Thomale2020B} proposed that NiO$_2$ terminated surface will harbor an extended $s$-wave gap as opposed to $d$-wave gap on Nd-terminated surface, primarily due to different local dopings. In presence of a step edge, Josephson coupling between the two regions will result in to $d+is$ state near the edge, leading to an evolution from V-shaped to U-shaped gap across the edge. One of the issues here is that the two surfaces are likely to have strong hybridization and not just a weak-link, which may lead to a uniform gap symmetry across the edges. Finally, using a two-orbital model, Ref. \cite{Das2020} obtained orbital-selective gaps with $d$-wave character on Ni-$d_{x^2 - y^2}$ orbital and extended $s$-wave gap on the axial orbital. Even though the gap has different symmetries in the orbital space, it must reduce to a unique symmetry in the band space, and hence, can not explain the STM observation as such. 

Here, we propose a simple explanation of the observed STM spectra using a two-band model \cite{Hepting2020} obtained by downfolding DFT results to Ni-$d_{x^2-y^2}$ and R-$d_{3z^2 - r^2}$ orbitals. Assuming $d$-wave pairing, we hypothesize that the superconducting gap follows a cuprate-like lowest-harmonic $d$-wave gap on Ni-$d_{x^2-y^2}$ band and {non-monotonic} $d$-wave gap with higher harmonics on R-$d_{3z^2 - r^2}$ band. We compute continuum local density of states (LDOS) measured by the STM tip at various intra-unit cell positions, utilizing a first-principles Wannier function-based approach \cite{Choubey2014, Kreisel2015, Choubey2017B} and find U-, V-, and mixed-shaped spectra, in agreement with the experiment. Very similar spectral line-shapes are also obtained for $d+is$-state and we propose a phase-sensitive quasiparticle interference (QPI) experiment \cite{HAEMS2015} to distinguish between the two. 

The paper is organized as follows. In Section \ref{sec:Model}, we introduce a mean-field model of superconducting {\RNO} with aforementioned $d$-wave gap structure, describe the procedure to compute continuum LDOS at intra-unit cell points, and construct a phase-senstive QPI observable \cite{HAEMS2015} to distinguish between $d$-wave and $d+is$-wave ground states. In Section \ref{sec:Results}, we present results detailing intra-unit cell spectra in nickelates, discuss the theoretical underpinnings of our numerical findings and comment on the sensitivity of our results on various parameters. Finally, we summarize our findings in Section \ref{sec:Conclusion}. 

\section{Model}
\label{sec:Model}
\begin{figure}[hbt!]
\begin{center}
\includegraphics[width=1\columnwidth]{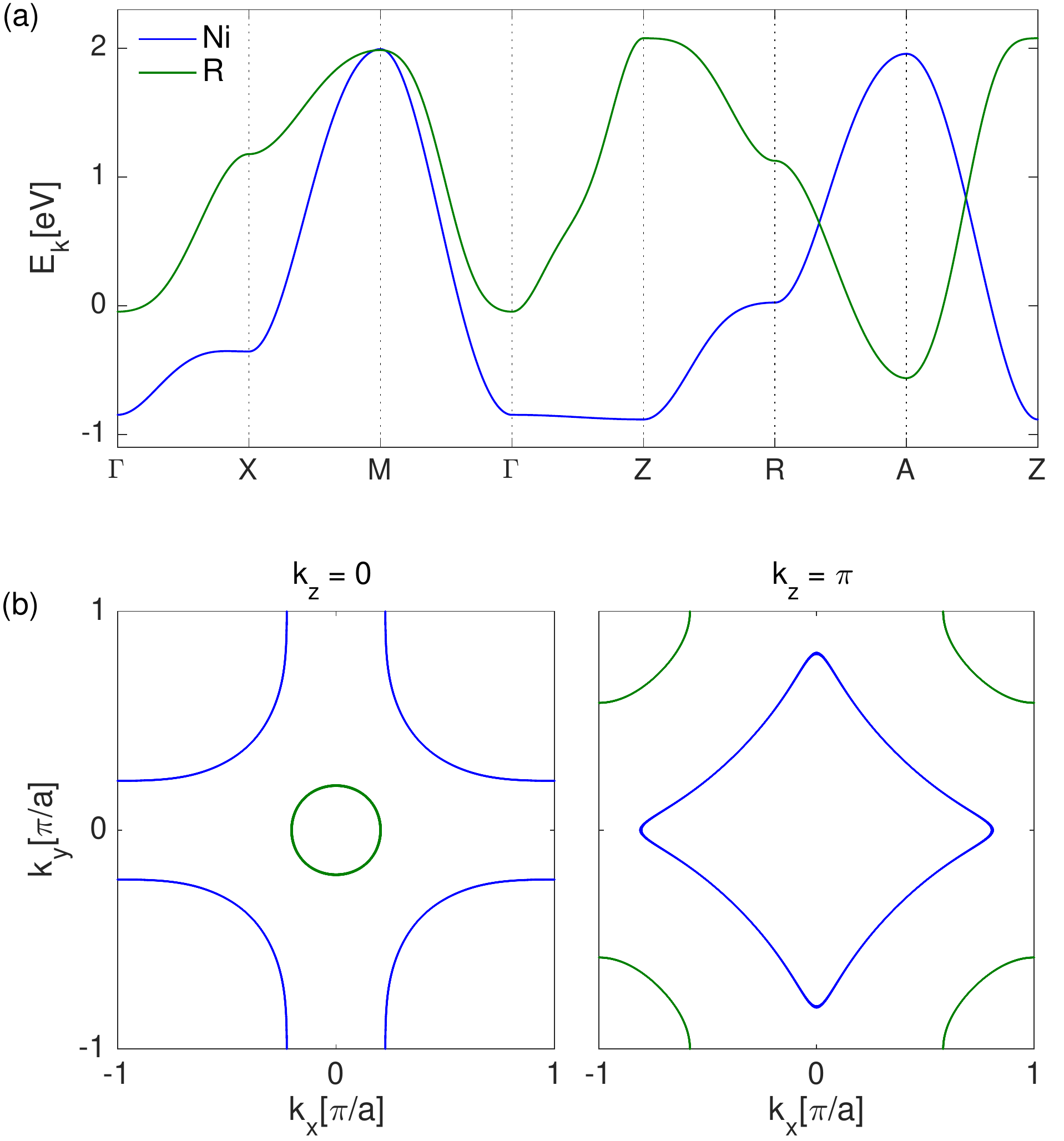}
\caption{Characteristic features of the two-band model proposed in Ref. \onlinecite{Hepting2020}. (a) Band structure of undoped {\RNO} (R = La) along $\Gamma$-$Z$ direction. Bands shown in green and blue has dominant Ni-$d_{x^2 - y^2}$ and R-$d_{3z^2 - r^2}$ character, respectively. $k_{z} = 0$ (b) and $k_{z} = \pi$ (c) cut of the Fermi surface of $15\%$ hole-doped {\RNO}, color coded according to the largest orbital weight.}
\label{fig:bands}
\end{center}
\end{figure}

\begin{figure}
\begin{center}
\includegraphics[width=1\columnwidth]{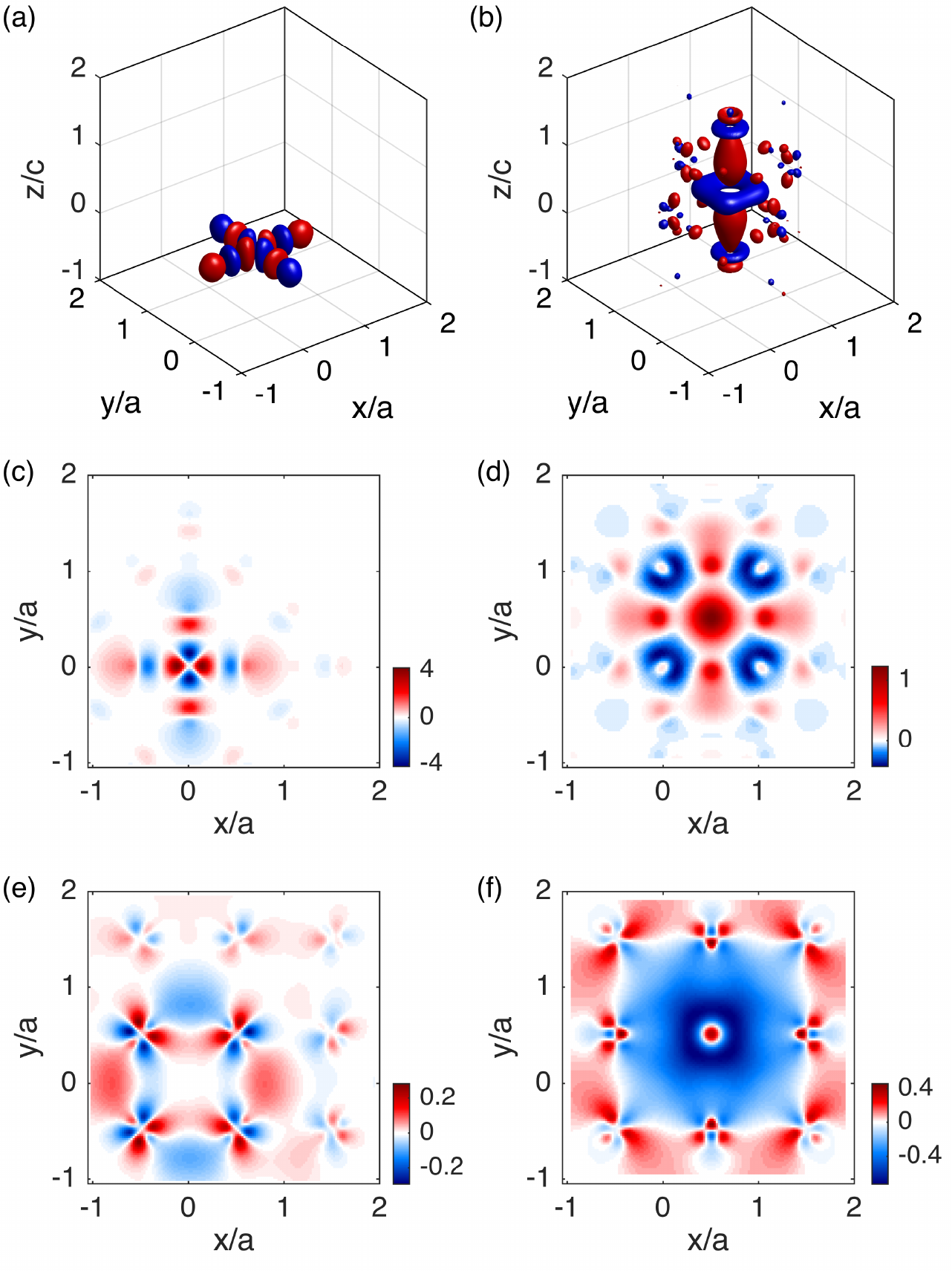}
\caption{Ni-$d_{x^2 - y^2}$ and R-$d_{3z^2 - r^2}$ Wannier function isosurfaces and z-cuts, {taken from Ref.\cite{Hepting2020}}. Isosurface plots of the Ni-$d_{x^2 - y^2}$ (a) and R-$d_{3z^2 - r^2}$ (b) Wannier function at  0.1\AA$^{-3/2}$. Ni-$d_{x^2 - y^2}$ Wannier function at height $z \approx 0.1c$ ($c = 3.89${\AA} is the $c$-axis lattice constant) above NiO- (c) and R-plane (d). R-$d_{3z^2 - r^2}$ Wannier function at height $z \approx 0.1c$ above NiO- (d) and R-plane (f). Colorbar values in (c)-(f) are in the units of \AA$^{-3/2}$.}
\label{fig:Wannier}
\end{center}
\end{figure}

We model the homogeneous superconducting state in nickelates by the following mean-field Hamiltonian
\begin{align}
&H = H_{0}+H_\text{SC}, \label{eq:H}\\
&H_{0} = \sum_{\K, \mu, \nu, \sigma} \epsilon_{\K}^{\mu\nu}c_{\K\mu\sigma}^{\dagger}c_{\K\nu\sigma} - \mu_{0}\sum_{\K, \mu, \sigma}c_{\K\mu\sigma}^{\dagger}c_{\K\mu\sigma}, \label{eq:H0}\\
%&= \sum_{\K, \alpha, \sigma} \xi_{\K\alpha}c_{\K\alpha\sigma}^{\dagger}c_{\K\alpha\sigma}, \label{eq:H0_band}\\
&H_\text{SC} = \sum_{\K, \alpha, \sigma} \Delta_{\alpha}(\K) c_{\K\alpha\uparrow}^{\dagger}c_{-\K\alpha\downarrow}^{\dagger} + \text{H.c.} \label{eq:H_SC}
\end{align}
In Eq.(\ref{eq:H0}), $c_{\K\mu\sigma}^{\dagger}$ creates an electron in Wannier orbital $\mu$ with spin $\sigma$ and momentum $\K$, $\epsilon_{\K}^{\mu\nu}$ is the normal-state dispersion in orbital basis, and $\mu_{0}$ is the chemical potential. In Eq. (\ref{eq:H_SC}), $c_{\K\alpha\sigma}$ creates an electron in band $\alpha$ with spin $\sigma$ and momentum $\K$, and $\Delta_{\alpha}(\K)$ is singlet superconducting gap on band $\alpha$. To describe the normal state of infinite-layer nickelates, we adopt a two-orbital model for {\RNO} (R = La) obtained by downfolding DFT-derived bands to Ni- and R-centered Wannier orbitals with {\dx} and {\dz}-symmetry, respectively \cite{Hepting2020}. The R-{\dz} derived band exhibits a three-dimensional (3D) dispersion whereas the Ni-{\dx} derived band exhibits a quasi-2D character, similar to cuprates, as evident from the band structure shown in Fig. \ref{fig:bands}(a). Ni-{\dx} band contributes to a large hole-like Fermi surface, whereas R-{\dz} band results in to a Fermi surface that changes from $\Gamma$-centered, small electron pocket in $k_{z} = 0$ plane to $A$-centered electron pocket in $k_{z} = \pi$ plane, see Fig. \ref{fig:bands}(b), (c). We note that the mixing between Ni-{\dx} and R-{\dz} orbital states is very small for a large energy window leading to the bands with essentially single-orbital characteristic\cite{Hepting2020}. Ni-{\dx} and R-{\dz} Wannier orbital's isosurfaces and 2D cuts obtained just above Ni- and R-plane are shown in Fig. \ref{fig:Wannier}. Ni-{\dx} Wannier orbital shows more localized and planar characteristics whereas R-{\dz} Wannier orbital is more extended and 3D in nature. Further, Ni-{\dx} orbital has vanishing weight above Ni site ($\rr_{\text{Ni}} = [0, 0, 0]$) as well as along R-Ni directions due to {\dx} symmetry, in contrast to R-\dz orbital, which has large weight above R site ($\rr_{\text{R}} = [0.5a, 0.5a, 0.5c]$, where $a$ and $c$ are lattice constants).

\begin{figure}
\begin{center}
\includegraphics[width=1\columnwidth]{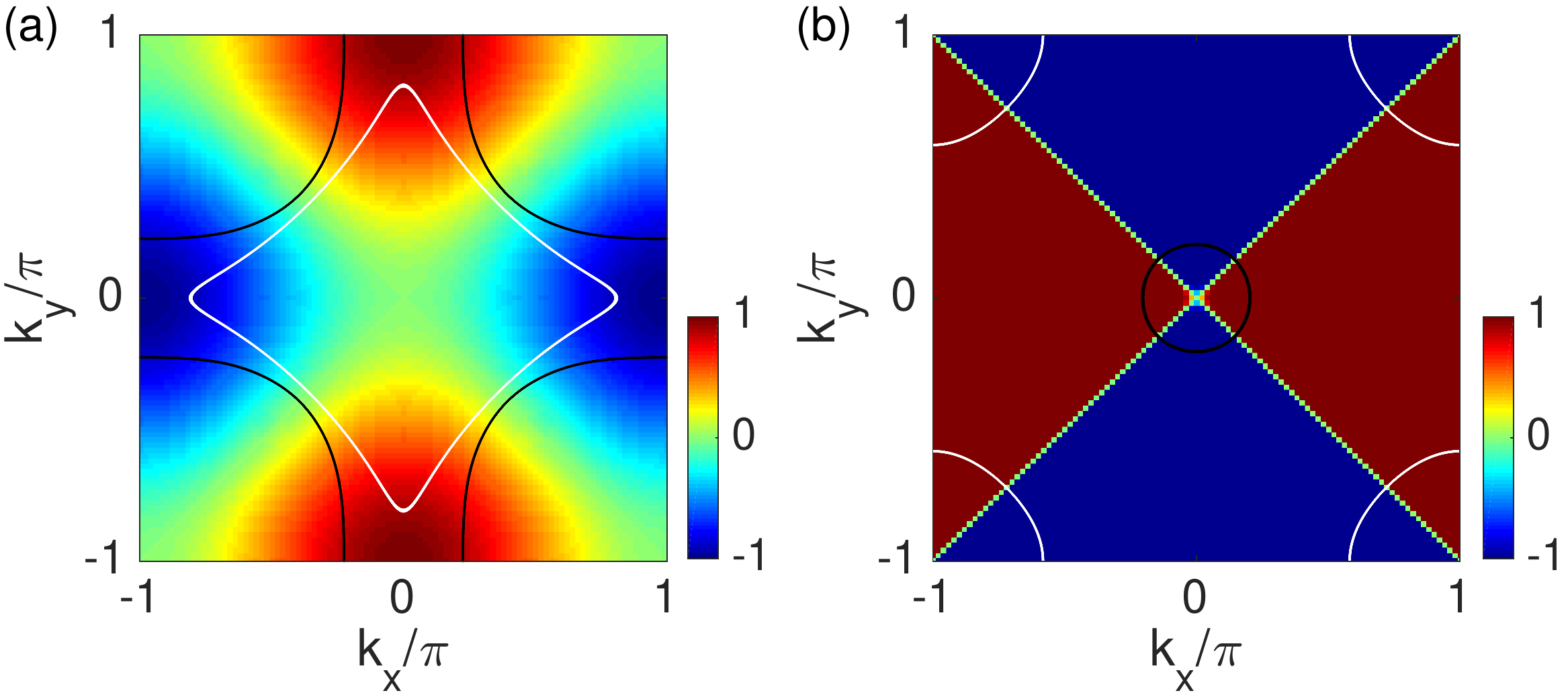}
\caption{Single-harmonic and higher-harmonic $d$-wave gap functions. (a) Conventional, single-harmonic, $d$-wave gap with form $cos(k_x) - cos(k_y)$, assumed to be present on Ni-$d_{x^2 - y^2}$ band. (b) Higher-harmonics $d$-wave gap function with form $\tanh{\alpha\left(k_{x}^2 - k_{y}^2\right)}$ ($\alpha = 100$), assumed to be present on R-$d_{3z^2 - r^2}$ band. In both plots, black and white solid lines depict $k_{z} = 0$ and $k_{z} = \pi$ Fermi surfaces, respectively, contributed by respective bands.}
\label{fig:gapFun}
\end{center}
\end{figure}

{Following previous theoretical proposals \cite{Arita2019, Kuroki2020, Thomale2020A, Das2020, Werner2020A}}, we assume that the superconducting pairing is mostly driven by Ni-{\dx} orbital, similar to hole-doped cuprates, leading to $d_{x^2-y^2}$-wave gap symmetry. Further, as the Ni-{\dx} band shows a dispersion very similar to Cu-{\dx} band in cuprates, it is reasonable to assume that the superconducting gap on the Ni-{\dx} band can be modeled by a simple lattice version of the $d_{x^2-y^2}$-wave gap structure as follows.
\begin{equation}
\label{eq:gap_Ni}
\begin{aligned}
\Delta_{\text{Ni}}(\K) = \frac{\Delta_{\text{Ni}}^{0}}{2}\left(\cos{k_{x}} - \cos{k_{y}}\right)
 \end{aligned}
\end{equation}
However, the R-\dz band shows significant dispersion along $k_{z}$, unlike quasi-2D Ni-{\dx} band. Although it is still expected that the superconducting gap on this band  possesses an overall $d_{x^2-y^2}$-wave symmetry mostly driven by the the interband Cooper-pair scattering to the dominant quasi-2D Ni-{\dx} band, it can naturally acquire a significant non-monotonic behavior  on the R-{\dz} band.  In particular, to explain the aforementioned STM observation \cite{Wen2020B}  and keeping the overall $d_{x^2-y^2}$-wave symmetry, we propose that the gap on R-{\dz} band has the following form.
\begin{equation}
\label{eq:gap_R}
\begin{aligned}
\Delta_{\text{R}}(\K) = \Delta_{\text{R}}^{0}\tanh{\left[\alpha\left(k_{x}^2 - k_{y}^2\right)\right]}
 \end{aligned}
\end{equation}
Fig. \ref{fig:gapFun} shows the gap structure in $k_{x}$-$k_{y}$ plane along with the $k_{z} = 0$ and $k_{z} = \pi$ Fermi surfaces derived from the R-{\dz} band. For a sufficiently large value of $\alpha$, the gap behaves essentially as a step function of the polar angle, sharply changing sign across the nodal directions while keeping a constant magnitude elsewhere. {This non-monotonic behavior of the superconducting gap is not unusual. For example, the non-monotonic $d_{x^2-y^2}$-wave gap is observed in electron-doped cuprates\cite{Blumberg2002,Matsui2005}, which is believed to be due to particular crossing of the AF Brillouin Zone boundary with the Fermi surface in these systems.
In the superconducting nickelates, the reason for the non-monotonic dependence of the $d_{x^2-y^2}$-wave gap on R-{\dz} band can be more isotropic magnetic response on this band as compared to the Ni-{\dx} band.} As we will discuss later, this particular gap structure is responsible for U-shaped partial LDOS contribution from the R-band. In addition, we will also consider a mixed $d+is$ gap structure \cite{Zhang2020B}, where single-harmonic $d$-wave gap (Eq. \ref{eq:gap_Ni}) exists on the Ni-band and an $s$-wave gap with $\pi/2$ phase difference exists on the R-band, i.e. $\Delta_{\text{R}}(\K) = i\Delta_{\text{R}}^{0}$.

By diagonalizing the normal state Hamiltonian, Eq. (\ref{eq:H0}), we can obtain total lattice LDOS $N_{\text{tot}}(\omega) = -\frac{2}{\pi}\text{Im}\left[\sum_{\K\alpha}G_{\K\alpha}(\omega)\right]$. Here, factor of $2$ accounts for the spin degeneracy, $\text{Im}$ represents imaginary part, and $G_{\K\alpha}(\omega)$ is the Greens function corresponding to band $\alpha$ which can be expressed as
\begin{equation}
\label{eq:GreensK_band}
\begin{aligned}
G_{\K\alpha}(\omega) = \frac{\vert u_{\K\alpha} \vert^2}{\omega - E_{\K\alpha} + i0^{+}} + \frac{\vert v_{\K\alpha} \vert^2}{\omega + E_{\K\alpha} + i0^{+}},
 \end{aligned}
\end{equation}
where, $\vert u_{\K\alpha} \vert^2 = \frac{1}{2}\left(1 + \frac{E_{\K\alpha}}{\xi_{\K\alpha}}\right)$ and $\vert v_{\K\alpha} \vert^2 = \frac{1}{2}\left(1 - \frac{E_{\K\alpha}}{\xi_{\K\alpha}}\right)$ are the Bogoliubov coherence factors, $ E_{\K\alpha} = \sqrt{\xi_{\K\alpha}^{2} + \vert \Delta_{\K\alpha} \vert^2}$ the quasiparticle energy in the superconducting state, $i0^{+}$ the artificial broadening, and $\xi_{\K\alpha}$ is the band dispersion in the normal state. Moreover, we can obtain the orbital-resolved LDOS $N^{\mu}(\omega) = -\frac{2}{\pi}\text{Im}\left[\sum_{\K}G_{\K}^{\mu\mu}(\omega)\right]$ using the orbital-space Greens function $G_{\K}^{\mu\nu}(\omega)$ that can be obtained from band-space Greens function $G_{\K\alpha}(\omega)$ (Eq. \ref{eq:GreensK_band}) via following band-to-orbital basis transformation.
\begin{equation}
\label{eq:GreensK_orbital}
\begin{aligned}
G_{\K}^{\mu\nu}(\omega) = \sum_{\alpha}U_{\alpha}^{\mu}(\K)G_{\K\alpha}(\omega)U_{\alpha}^{\nu\ast}(\K).
 \end{aligned}
\end{equation}
Here, $U_{\alpha}^{\mu}$ are the elements of $2\times2$ unitary matrix $U$ that diagonalizes the normal state Hamiltonian $H_{0} = \sum_{\K\mu\nu\sigma}c_{\K\mu\sigma}H_{0}^{\mu\nu}(\K)c_{\K\nu\sigma}$, with matrix elements $H_{0}^{\mu\nu}(\K) = \epsilon^{\mu\nu}_{\K} - \mu_{0}\delta_{\mu\nu}$, yielding the normal-state band dispersion $\xi_{\K\alpha}$. 

The differential conductance measured in STM experiments is proportional to the LDOS evaluated at the position of the STM tip \cite{Tersoff1985}, and hence, orbital-resolved (or total) lattice LDOS are not yet the appropriate quantities to compare with experiment as we need to find LDOS at continuum positions $\rr$ traversed by the STM tip. This can be achieved by a recently developed first-principles Wannier function based approach \cite{Choubey2014, Kreisel2015, Choubey2017B}, which has been applied successfully to interpret and predict various STM observables in cuprates \cite{Kreisel2015, Choubey2017A, Choubey2017B, Choubey2020, Boeker2020, Wang2021} and Fe-based superconductors \cite{Choubey2014, Kreisel2016, Chi2016}. Here, continuum LDOS at the STM tip position $\rr$ is given by $\rho(\rr, \omega) = -\frac{2}{\pi}\text{Im}\left[G\left(\rr, \omega\right)\right]$, where, the Greens function $G(\rr, \omega)$ defined in continuum space is related to the usual lattice Greens function $G_{ij}^{\mu\nu}(\omega)$ that connects orbital $\mu$ at site $i$ to orbital $\nu$ at site $j$ via a basis transformation, where Wannier functions $W_{i}^{\mu}(\rr)$ serve as the matrix elements of the transformation:
\begin{equation}
\label{eq:GreensR_lattice}
\begin{aligned}
G\left(\rr, \omega\right) = \sum_{ij\mu\nu}W_{i}^{\mu}(\rr)G_{ij}^{\mu\nu}(\omega)W_{j}^{\nu\ast}(\rr).
 \end{aligned}
\end{equation}
Taking Fourier transform to $\K$-space and using Eq. \ref{eq:GreensK_band}, we obtain 
\begin{align}
&G\left(\rr, \omega\right) = \sum_{\K\alpha}G_{\K\alpha}(\omega)\vert W_{\K\alpha}(\rr) \vert^2, \label{eq:GreensR_kSpace}\\
&W_{\K\alpha}(\rr) = \sum_{\mu} U_{\alpha}^{\mu}(\K)W_{\K}^{\mu}(\rr), \label{eq:WannierK_band}\\
&W_{\K}^{\mu}(\rr) = \sum_{i} W_{i}^{\mu}(\rr)e^{i\K\cdot\R_{i}}. \label{eq:WannierK_orbital}
\end{align}
Finally, using Eqs. (\ref{eq:GreensR_kSpace})- (\ref{eq:WannierK_orbital}), we can obtain LDOS at any intra-unit cell point with a spatial resolution that is limited only by the resolution of the Wannier functions. In what follows, we use this approach to compute LDOS in $x-y$ planes located at heights slightly above Ni and R atoms; obtaining U, V, and mixed spectral line-shapes depending on the location in the planes.

\section{Results and Discussion}
\label{sec:Results}

\begin{figure}
\begin{center}
\includegraphics[width=1\columnwidth]{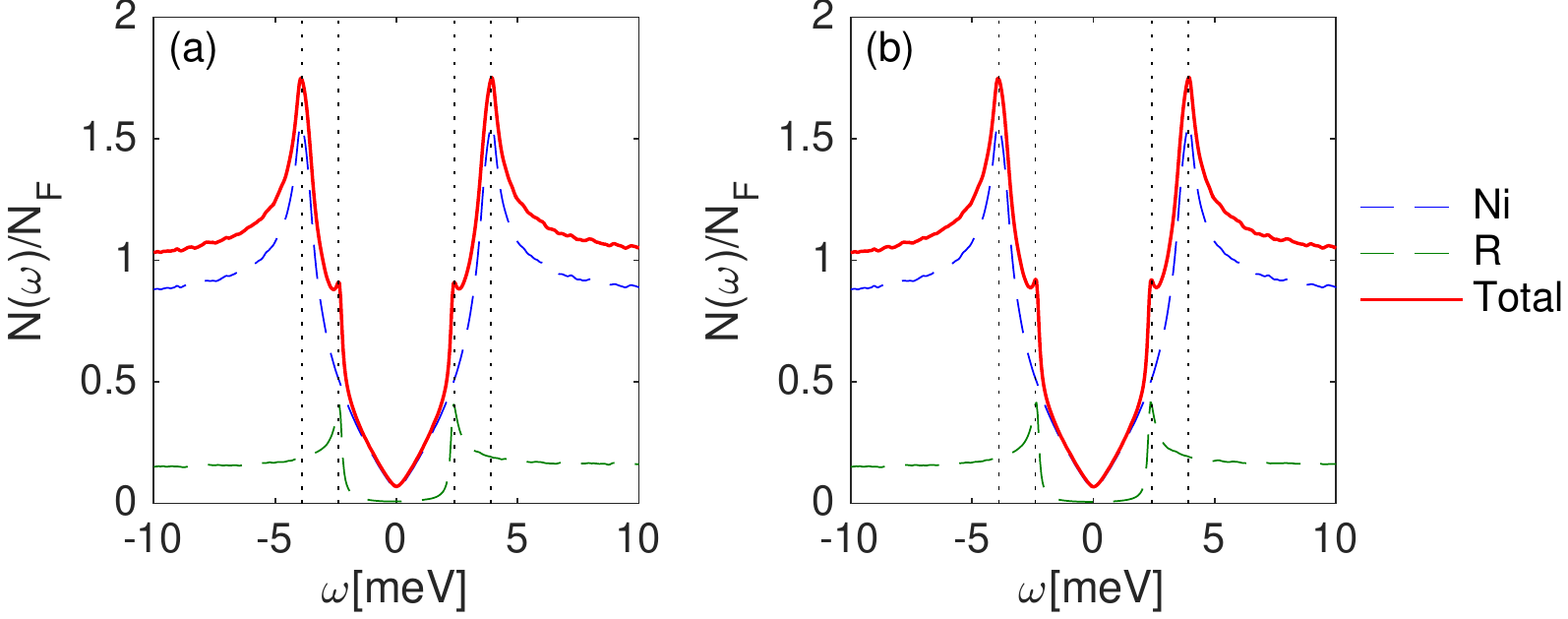}
\caption{Lattice density of states spectra. Orbital-resolved lattice LDOS in the proposed $d$-wave (a) and $d+is$-wave state, normalized with respect to the Fermi surface DOS in the normal state. Dotted vertical lines indicate two sets of coherence peaks at $\omega = \pm3.9$meV and $\omega = \pm2.35$meV, contributed by Ni-$d_{x^2 - y^2}$ and R-$d_{3z^2 - r^2}$ bands, respectively.}
\label{fig:lattice_LDOS}
\end{center}
\end{figure}

 We start discussing our numerical results by showing in Fig. \ref{fig:lattice_LDOS}(a) the  orbital-resolved lattice LDOS (normalized by the LDOS in the normal state) in the two-band model of RNiO$_2$, Eq. \ref{eq:H}, with simple $d$-wave gap on Ni-band (Eq. \ref{eq:gap_Ni}) and higher-harmonic $d$-wave gap on R-band (Eq. \ref{eq:gap_R}). Here, we have set the higher-harmonic gap parameter $\alpha = 100$, $\Delta_{\text{Ni}}^{0} = 4$meV, and $\Delta_{\text{R}}^{0} = 2.3$meV to obtain experimentally reported \cite{Wen2020B} two-gap features at $\pm3.9$mev and $\pm2.3$mev. The Ni-LDOS shows usual V-shaped spectrum expected from a simple $d$-wave gap, whereas the R-LDOS shows a U-shaped spectrum owing to the higher-harmonic gap structure. In fact, with increasing $\alpha$, the transition between positive and negative values of the superconducting gap on R-band becomes sharper leading to a decrease in the magnitude of the slope of R-LDOS at $\omega = 0$, which becomes vanishingly small at sufficiently large values of $\alpha$ (see Appendix \ref{appendix_UshapedLDOS}), resulting in a $s$-wave-like U-shaped gap on R-band. The $d+is$ gap structure yields a very similar orbital-resolved lattice LDOS, see Fig. \ref{fig:lattice_LDOS}(b). Thus, at the level of lattice LDOS, we can explain the experimentally observed V- and U-shaped gap as orbital-resolved gaps at Ni and R sites, respectively, emerging either from higher-harmonic $d$-wave or $d+is$ gap structure. However, the occurrence of experimentally observed U-V mixed-shaped spectra still remains a puzzle, which can be resolved only when we consider continuum LDOS.
 \begin{figure*}
\begin{center}
\includegraphics[width=2\columnwidth]{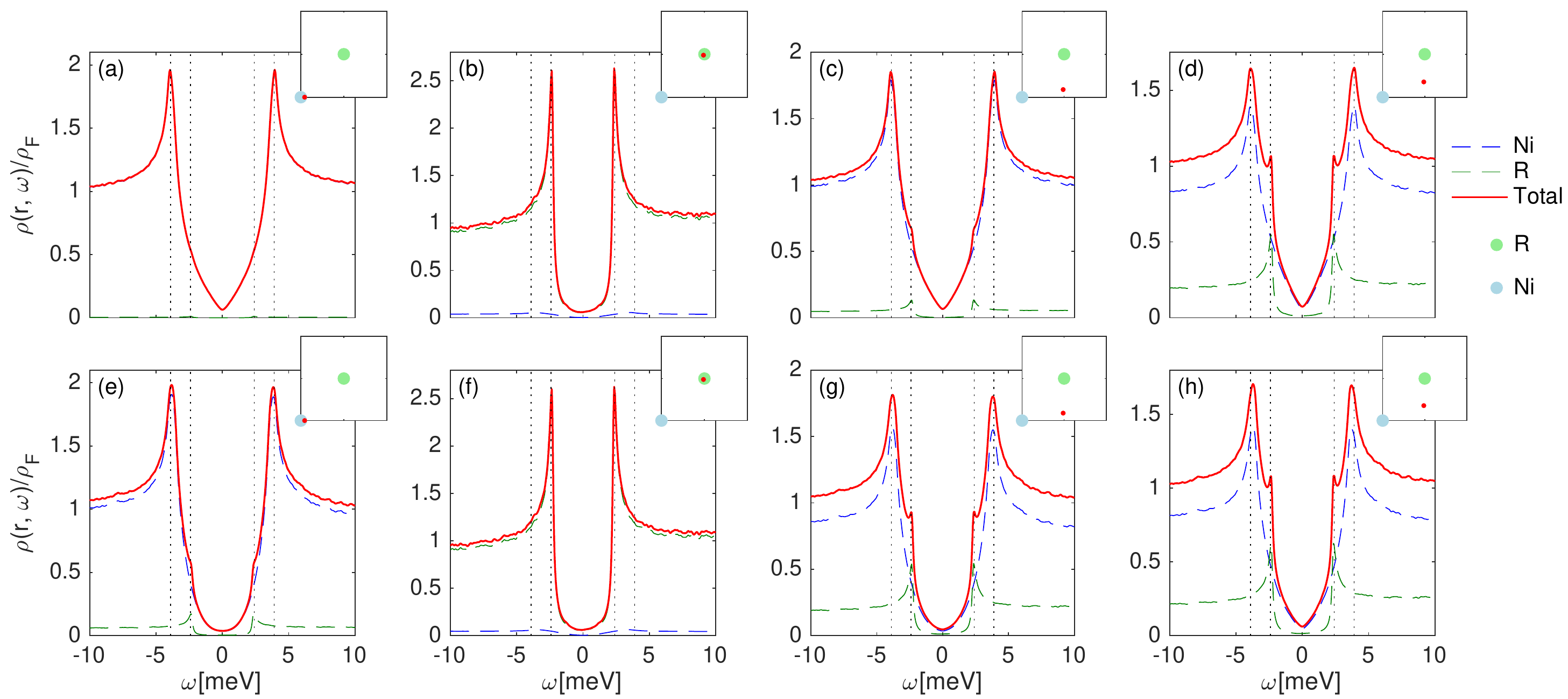}
\caption{Distinct spectral line shapes displayed by continuum LDOS in higher-harmonics $d$-wave state. Band-resolved continuum LDOS in the proposed $d$-wave state (normalized with respect to the Fermi surface LDOS) at a number of intra-unit cell points in the x-y plane located at a height $z \approx 0.1c$ above NiO-plane (a)-(d) and R-plane (e)-(h). Contributions to total LDOS (red curves) by Ni- and R-bands are plotted in blue and green, respectively. In each panel, inset shows the position of the intra-unit cell point (red dot) relative to Ni (filled circle in light blue) and R (filled circle in light green) positions. Dotted vertical lines indicate two sets of coherence peaks at $\omega = \pm3.9$meV and $\omega = \pm2.35$meV.}
\label{fig:cLDOS}
\end{center}
\end{figure*}

\begin{figure*}
\begin{center}
\includegraphics[width=1.5\columnwidth]{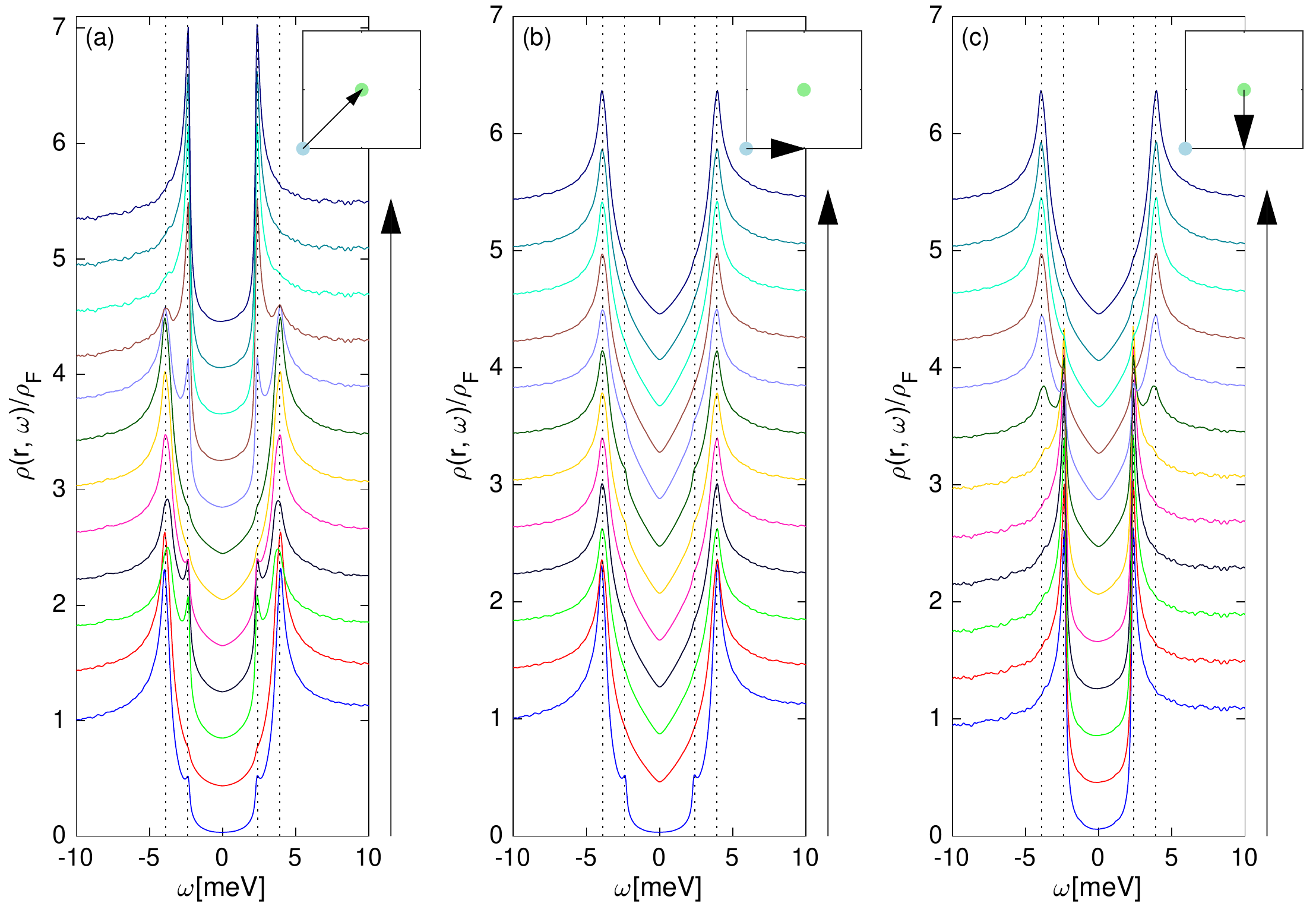}
\caption{Spectral line-cuts in higher-harmonics $d$-wave state along Ni-R (a), Ni-O (b), and R-O (c) directions in the $x$-$y$ plane located at a height $z \approx 0.1c$ above NiO-plane. Each spectrum is offset by 0.4 for clarity. Black arrows in insets indicate the cut directions with respect to Ni- (light-blue filled circle) and R- (light-green filled circle) positions in the $x$-$y$ plane. Dotted vertical lines indicate two sets of coherence peaks at $\omega = \pm3.9$meV and $\omega = \pm2.35$meV.}
\label{fig:lineCuts}
\end{center}
\end{figure*}

As explained in Section \ref{sec:Model}, the correct physical quantity which one should consider while theoretically interpreting STM results is continuum LDOS evaluated at the STM tip position (and not the lattice LDOS). Figure \ref{fig:cLDOS} shows total continuum LDOS along with contributions from Ni- and R-band at various intra-unit cell points located in planes right above Ni- and R-planes, calculated using Eq. \ref{eq:GreensR_kSpace} utilizing Ni-$d_{x^2 - y^2}$ and R-$d_{3z^2 - r^2}$ Wannier functions \cite{Hepting2020} described in Section \ref{sec:Model}. The LDOS spectra show three distinct lineshapes: U-shaped, V-shaped, and mixed-shaped, depending on the position of the intra-unit cell point. The origin of these distinct lineshapes can be understood as follows. Owing to the negligible mixing between the two bands ($U_{\alpha}^{\mu}(\K) \approx \delta_{\alpha\mu}$) for the energies of our interest, the k-space Wannier function in band space ($W_{\K\alpha}(\rr)$) (Eq. \ref{eq:WannierK_orbital}) and that in the orbital space ($W_{\K}^{\mu}(\rr)$) are almost identical. Further, the k-space Greens function in band space is approximately diagonal and equal to the orbital space Greens function. With these simplifications, the continuum Greens function (Eq. \ref{eq:GreensR_kSpace}) turns out to be a weighted sum of $k$-space Greens function with weighting factors given by $k$-space Wannier functions: $G\left(\rr, \omega\right) \approx \sum_{\K}G_{\K}^{\mu\mu}(\omega)|W_{\K}^{\mu}(\rr)|^{2}$, where $\mu = $ Ni-$d_{x^2 - y^2}$ and R-$d_{3z^2 - r^2}$. Consequently, we obtain a U(V)-shaped LDOS emerging from R(Ni)-band at the intra-unit cell points $\rr$ where Ni-$d_{x^2 - y^2}$ (R-$d_{3z^2 - r^2}$) Wannier function's contribution is negligible. A U-V mixed lineshape is obtained at the points where contributions from the both Wannier functions are comparable. Our results are consistent with recent STM experiments \cite{Wen2020B} and show that the mixture of U and V-shaped gap can be still a signature of the global $d$-wave gap. {{A more detailed STM investigation on a high quality sample is required to further test our hypothesis. The sample used in Ref. \onlinecite{Wen2020B} exhibited large surface roughness (of order 1nm)} making it impossible to ascertain the precise location of the STM tip relative to Ni/R-atoms. If such assignments can be made in future experiments then our theory can be easily tested. Fig. \ref{fig:lineCuts} shows the spectra obtained along three directions namely Ni-R, Ni-O, and R-O, in a plane right above the NiO plane. We find that in the Ni-O direction (Fig. \ref{fig:lineCuts}(a)), the spectral lineshapes change from mixed U-V with both coherence peaks (at $\pm$2.35meV and $\pm$3.9meV) to U-shaped gap with coherence peaks only at the lower gap edge (at $\pm$2.35meV). In contrast, the lineshpaes change from a U-shape (with coherence peaks at the lower gap edge) to V-shape (with coherence peaks at the higher gap edge) via mixed U-V shape (with coherence peaks at both gap edges) along the R-O direction (Fig. \ref{fig:lineCuts}(c))}. 

\begin{figure}
\begin{center}
\includegraphics[width=1\columnwidth]{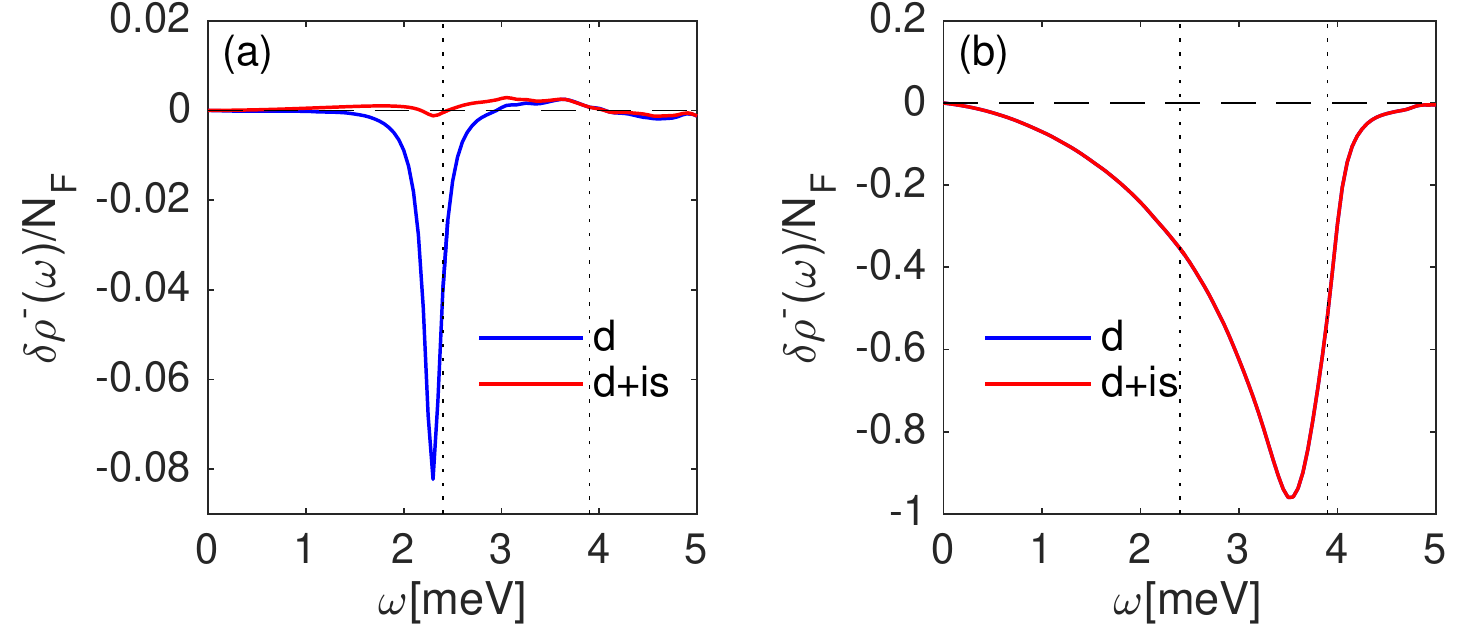}
\caption{HAEMS signature of the proposed $d$-wave and $d+is$-wave states. Antisymmetrized correction to the lattice LDOS due to scattering off a weak point-like impurity substituting R atom (a) and Ni atom (b) in the proposed $d$-wave state and $d+is$-wave state. Dotted vertical lines indicate positive coherence peaks at $\omega = 3.9$meV and $\omega = 2.35$meV. }
\label{fig:HAEMS}
\end{center}
\end{figure}

{{The continuum LDOS lineshapes at a given intra-unit cell point are very similar for both higher-harmonic $d$-wave gap model and $d+is$-gap model. % (see Figure \ref{fig:cLDOS_d_is} in Appendix \ref{appendix_moreDetailsSpectra}).
Thus, by itself the experimental data\cite{Wen2020B} cannot rule out the global $d$-wave gap in infinite layer nickelates, but instead call for further investigations.}} We recall that the two scenarios can be distinguished by a method proposed by Hirschfeld, Altenfeld, Eremin, and Mazin (HAEM) \cite{HAEMS2015}, which is designed to determine the sign structure of superconducting gap in a multi-band system by utilizing the quasiparticle interference (QPI) information from STM measurements. Following the HAEM's prescription, we construct antisymmetrized impurity-induced LDOS $\delta\rho^{-}(\omega) = \delta\rho(\omega) - \delta\rho(-\omega)$ using T-matrix formalism, see Appendix \ref{appendix_HAEM} for details. $\delta\rho^{-}(\omega)$ shows different behaviour depending on whether the impurity substitutes Ni site or R site. In the former case, it shows identical behavior for both higher-harmonic $d$-wave and $d+is$-wave gaps, attaining the minimum value near the second gap edge (at 3.9meV) without any change of sign, see Fig. \ref{fig:HAEMS}(b). A very similar behaviour in case of single-band cuprates was observed in Ref. \cite{Boeker2020}. However, in the latter case where an R-site is substituted by an impurity, $\delta\rho^{-}(\omega)$ changes sign and has much smaller magnitude in $d+is$ state, whereas it shows a minimum at the lower gap edge (2.35meV) without any sign-change for higher-harmonic $d$-wave state, see Fig. \ref{fig:HAEMS}(a). Therefore, the main difference in the HAEM signal occurs for the impurity on the R-site due to a very different phase structure of the gap on the R-band in both cases. This is the scenario which can be used to distinguish between the two gap structures.  % Ho show identical behavior  As Figure \ref{fig:HAEMS}(a) shows,  Originally envisaged to distinguish between sign-preserving ($s_{++}$) and sign-changing ($s_{+-}$) s-wave superconducting gap on two bands, HAEM's method involves construction of antisymmetrized impurity-induced LDOS $\delta\rho^{-}(\omega) = \delta\rho(\omega) - \delta\rho(-\omega)$ which shows very different behavior in the two cases. In $s_{++}$

\section{Conclusion}
\label{sec:Conclusion}
To conclude, we have presented a theoretical analysis of the U-, V-, and mixed-shape spectra observed in a recent STM study on superconducting {\NNO} thin films assuming a global $d$-wave symmetry of the superconducting state. Starting with a recently proposed two-orbital model for the normal state of {\RNO} (R = La), which yields a quasi-2D band with Ni-{\dx} character (very similar to cuprates) and a 3D band with R-{\dz} character exhibiting strong dispersion in $\Gamma-Z$ direction, we assumed that a global $d$-wave symmetric superconducting state is driven by the Ni $3d_{x^2-y^2}$-band resulting in a simple $(\cos{k_{x}} - \cos{k_{y}})$ form of the superconducting gap on the same, mirroring hole-doped cuprates. Further, we conjectured that the superconducting gap on the R-band consisted of higher $d$-wave harmonics that can be represented by $\tanh(\alpha(k_{x}^2 - k_{y}^2))$ form. This results in V-shaped lattice LDOS at the Ni site and, for sufficiently large $\alpha$, U-shaped LDOS at the R-site. The correct quantity to compare with the differential conductance spectra measured by STM is the continuum LDOS, however, which we computed using first-principles Ni-{\dx} and R-{\dz} Wannier functions reported in Ref. \onlinecite{Hepting2020}. We found that depending on intra-unit cell position; U-, V-, and mixed-shape spectrum can be obtained within a {\RNO} unit cell. In essence, the $k$-space Wannier functions act as weighting factors for Green's functions contributed by Ni- and R-bands, yielding U-shaped (V-shaped) continuum LDOS at the intra-unit cell positions where contributions from R-{\dz} (Ni-{\dx}) Wannier function dominates. A mixed-shape spectra is obtained at the positions where both Wannier functions have comparable contributions. 

The scenario we proposed can be tested in future STM experiments with improved sample quality where position of the STM tip with respect to the {\RNO} unit cell can be precisely determined. In particular, line-scans along Ni-R and R-O directions should yield spectral lineshapes evolving from mixed- to U-shaped and U- to V-shaped, respectively. Further, we showed that a $d + is$-gap structure yields spectra nearly identical to the higher-harmonic $d$-wave gap model and the two scenarios can be distinguished by phase-sensitive QPI measurements on {\RNO} samples with R-sites substituted by impurities. Finally, we note that further work is required to understand the genesis of the higher-harmonic $d$-wave gap from a microscopic theory, which we intend to pursue in future.  

\section{Acknowledgements}
The authors wish to acknowledge useful discussions with F. Lechermann, H. H. Wen, F. Jakubczik, A. R\"omer, and B. M. Andersen. The authors are grateful to T. P. Devereaux and Chunjing Jia for providing the Wannier function data published in Ref. \onlinecite{Hepting2020}.  The work was supported by the joint NSFC-DFG grant (ER 463/14-1)

\appendix
%\section{Two-band model}
%\label{appendix_twoBandModel}

\section{U-shaped lattice LDOS in higher-harmonic $d$-wave state}
 \label{appendix_UshapedLDOS}
 \renewcommand\thefigure{\thesection.\arabic{figure}}  
 \setcounter{figure}{0} 
In this appendix, we will show that the higher-harmonic $d$-wave gap function in Eq. \ref{eq:gap_R} leads to a lattice LDOS that varies linearly with bias at low-energies: $N(\omega) = \gamma |\omega|$ for $\omega \rightarrow 0$, with slope $\gamma \propto 1/\alpha$, which becomes very small for for sufficiently large values of the higher-harmonic gap parameter $\alpha$, resulting in a U-shaped spectrum. We begin with the spectral function in the superconducting state:
\begin{equation}
\label{eq:spectral_function}
\begin{aligned}
A_{\sigma}(\mathbf{k},\omega) = \vert u_{\mathbf{k}}\vert^2 \delta(\omega - E_{\mathbf{k}}) + \vert v_{\mathbf{k}}\vert^2 \delta(\omega + E_{\mathbf{k}}),
 \end{aligned}
\end{equation}
where, coherence factors and quasiparticle energy are given by 
\begin{align}
&\vert u_{\mathbf{k}}\vert^2 = \frac{1}{2}\left(1 + \frac{\xi_{\mathbf{k}}}{E_{\mathbf{k}}}\right) = 1 - \vert v_{\mathbf{k}}\vert^2, \label{eq:ukvk}\\
&E_{\mathbf{k}} = \sqrt[]{\xi_{\mathbf{k}}^2 + \Delta_{\mathbf{k}}^2} \label{eq:Ek},
\end{align}
with normal state dispersion $\xi_{\K}$ and gap function $\Delta_{\mathbf{k}}$, which is given by higher-harmonic $d$-wave form in Eq. \ref{eq:gap_R}. In what follows, we work in the units where $\hbar = 1$, suppress the spin index $\sigma$, and set the lattice constant $a = 1$ so that the higher-harmonic gap coefficient $\alpha$ can be taken as a dimensionless constant. For $\omega > 0$, the lattice LDOS can be expressed as
\begin{equation}
\label{eq:latticeDOS}
\begin{aligned}
N(\omega) = 2\sum_{\K}A(\mathbf{k},\omega) = \int \frac{d\K}{(2\pi)^2}\left(1 + \frac{\xi_{\mathbf{k}}}{E_{\mathbf{k}}}\right)\delta(\omega - E_{\mathbf{k}}).
 \end{aligned}
\end{equation}
At low-energies, dominant contributions to the integral in Eq. \ref{eq:latticeDOS} will be from nodal regions. Accordingly, we can approximate the integral as
\begin{equation}
\label{eq:latticeDOS_approx}
\begin{aligned}
N(\omega\rightarrow 0) \approx M\int_{\Omega} \frac{d\K}{(2\pi)^2}\left(1 + \frac{\xi_{\mathbf{k}}}{E_{\mathbf{k}}}\right)\delta(\omega - E_{\mathbf{k}}),
 \end{aligned}
\end{equation}
where, $M = 4$ is the number of nodes and $\Omega$ is a  small circular region of radius $\Gamma$ around a nodal point ${\K}_{0}$, see Fig. \ref{fig:nodal_coordinates}. Further, to facilitate the computation of the integral, we introduce new coordinate system with origin at $\K_{0}$ and axes $k_{1}$ and $k_{2}$ along and perpendicular to the corresponding nodal direction, respectively, see Fig. \ref{fig:nodal_coordinates}. 
%\begin{equation}
%\begin{aligned}
%&k_{1} = \frac{1}{\sqrt[]{2}}(k_{y} + k_{x}) - k_{0}\\
%&k_{2} = \frac{1}{\sqrt[]{2}}(k_{y} - k_{x}).
%\end{aligned}
%\label{eq:k1k2}
%\end{equation}
Furthermore, we expand $\xi_{\K}$ and $\Delta_{\mathbf{k}}$ around the node up to linear order. The result can be expressed in the new coordinate system as
\begin{equation}
\label{eq:linearized_xiK}
\begin{aligned}
\xi_{\mathbf{k}} \approx v_{F}k_{1},
 \end{aligned}
\end{equation}
\begin{equation}
\label{eq:linearized_deltaK}
\begin{aligned}
\Delta_{\mathbf{k}} \approx -v_{\Delta}k_{2},
 \end{aligned}
\end{equation}
where, $v_{F}$ is Fermi velocity at the node and %$v_{\Delta}$ is given by
\begin{equation}
\label{eq:vDelta}
\begin{aligned}
v_{\Delta} = 2\alpha\Delta_{0}k_{0}.
 \end{aligned}
\end{equation}
We note that in case of the simple $d$-wave gap, $v_{\Delta} = \sqrt{2}\Delta_{0}\sin{\left(k_{0}/\sqrt{2}\right)}$. Now, first substituting Eq. \ref{eq:linearized_xiK} and \ref{eq:linearized_deltaK} in Eq. \ref{eq:latticeDOS_approx}, then scaling the coordinates: $k_{1}^{\prime} = v_{F}k_{1}$, $k_{2}^{\prime} = v_{\Delta}k_{2}$, and finally moving to the polar coordinates: $k_{1}^{\prime} = k^{\prime}\cos{\theta^{\prime}}$, $k_{2}^{\prime} = k^{\prime}\sin{\theta^{\prime}}$, we obtain
\begin{align}
&N(\omega) = \frac{M}{(2\pi)^2}\frac{1}{v_{F}v_{\Delta}}\left(I_{1} + I_{2}\right), \\
&I_{1} = \int_{0}^{2\pi}d\theta^{\prime}\int_{0}^{\Gamma}k^{\prime}dk^{\prime}\delta(\omega - k^{\prime}),\\
&I_{2} = \int_{0}^{2\pi}d\theta^{\prime}\cos{\theta^{\prime}}\int_{0}^{\Gamma}k^{\prime}dk^{\prime}\delta(\omega - k^{\prime}). %\label{eq:H_SC}
\end{align}
Clearly, $I_{1} = 2\pi\omega$ and $I_{2} = 0$, thus the low-energy LDOS for $\omega>0$ is given by 
\begin{equation}
\label{eq:DOSplus}
\begin{aligned}
N(\omega) = \left(\frac{M}{2\pi v_{F}v_{\Delta}}\right)\omega,
 \end{aligned}
\end{equation}
Similarly, we can obtain the LDOS for $\omega < 0$. Combining both together, we get
\begin{equation}
\label{eq:DOS}
\begin{aligned}
N(\omega) =\left(\frac{1}{\pi v_{F}\Delta_{0}k_{0}\alpha}\right)|\omega|,
 \end{aligned}
\end{equation}
where, we have used Eq. \ref{eq:vDelta} and $M=4$. Thus, we see that the lattice LDOS at low-energies vary linearly with bias with slope $\gamma = 1/\left(\pi v_{F}\Delta_{0}k_{0}\alpha\right)$, which can be very small at sufficiently large values of $\alpha$ leading to a U-shaped spectrum. 

\begin{figure}
\begin{center}
\includegraphics[width=0.7\columnwidth]{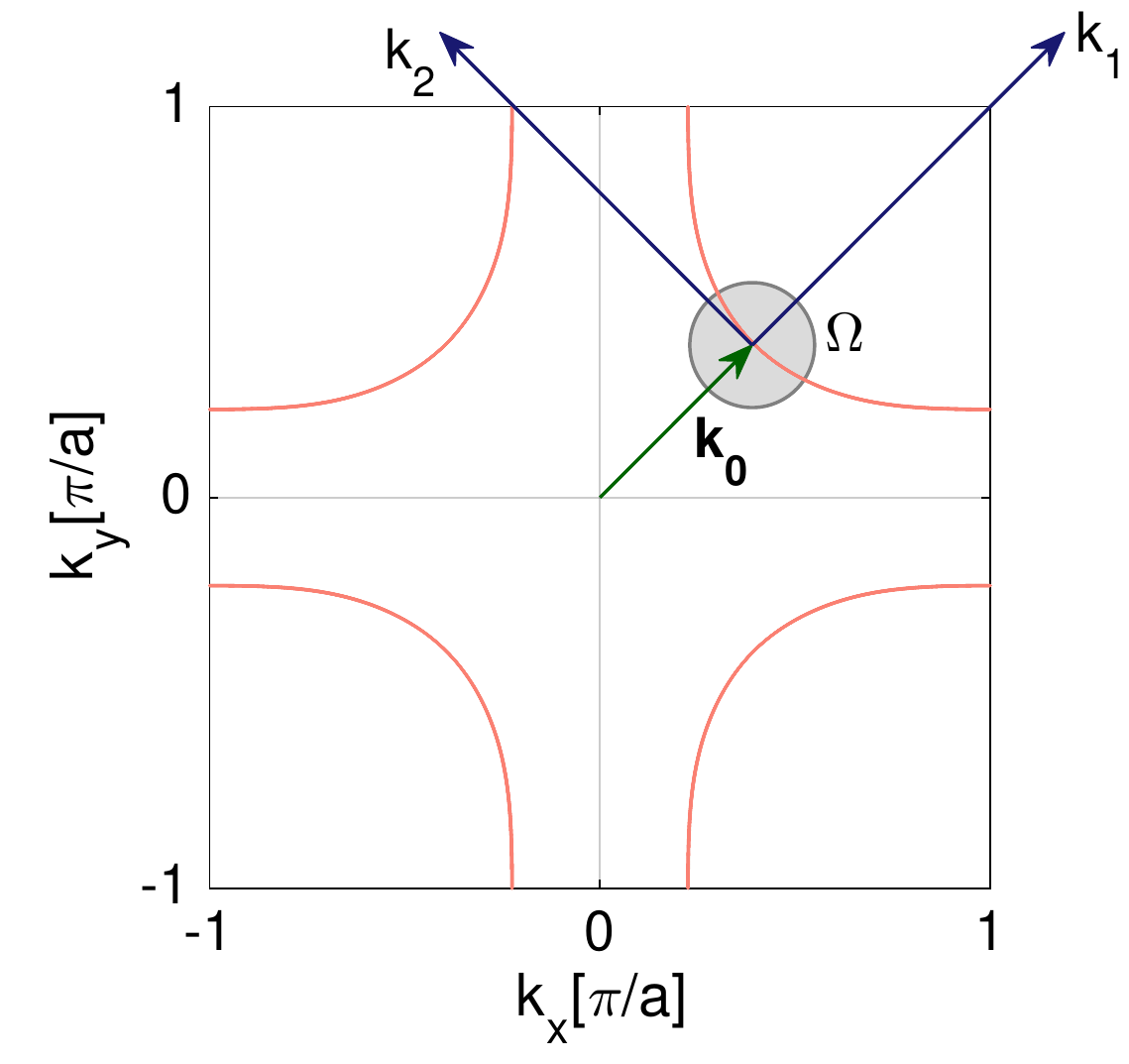}
\caption{Schematic of a representative Fermi surface superimposed on a nodal coordinate system with origin at the nodal point $\K_{0}$ (green arrow) and axes $k_1$ and $k_2$ parallel and perpendicular to the nodal direction, respectively. The region of integration $\Omega$ in Eq. \ref{eq:latticeDOS_approx} is shown by gray filled circle around the node.} 
\label{fig:nodal_coordinates}
\end{center}
\end{figure}

%\section{More details on intra-unit cell spectra}
% \label{appendix_moreDetailsSpectra}

 \section{Calculation of anti-symmetrized correction to LDOS due to impurity scattering}
 \label{appendix_HAEM}
This appendix provides a detailed description of the T-matrix formalism as applied to the construction of anti-symmetrized impurity-induced LDOS $\delta\rho^{-}(\omega) = \delta\rho(\omega) - \delta\rho(-\omega)$ (Fig. \ref{fig:HAEMS}) used in HAEM's scheme \cite{HAEMS2015}. We begin with the calculation of local Green's function $\hat{G}_{0}(\omega) = \sum_{\K}\hat{G}_{\K}(\omega)$, where $\hat{G}$ is the matrix Green's function in the orbital space with elements given by Eq. \ref{eq:GreensK_orbital}. Now, $T$-matrix for scattering off a point-like impurity substituting a lattice site (Ni or R) in the {\RNO} unit cell will be given by
\begin{equation}
\label{eq:Tmatrix}
\begin{aligned}
\hat{T}(\omega) = \left[\mathbb{1} - \hat{V}\hat{G}_{0}(\omega)\right]^{-1}\hat{V}.
 \end{aligned}
\end{equation}
Here, $\mathbb{1}$ represents identity matrix and $\hat{V} = \tau_{3}\otimes diag([V_{\text{imp}}^{Ni}, V_{\text{imp}}^{R}])$ is the impurity potential in Nambu space, where $\tau_{i}$ is a Pauli matrix, $diag$ represents a diagonal matrix, and $V_{\text{imp}}^{Ni (R)}$ is the on-site scattering potential for the impurity substituting an Ni (R) site. Impurity-induced change in the local Green's function can be obtained as
\begin{equation}
\label{eq:delG}
\begin{aligned}
\delta\hat{G}(\omega) = \hat{G}_{0}(\omega)\hat{T}(\omega)\hat{G}_{0}(\omega).
 \end{aligned}
\end{equation}
The corresponding (wavevector integrated) change in the total LDOS will be given by the following expression \cite{Boeker2020}:
\begin{equation}
\label{eq:delRho}
\begin{aligned}
\delta\rho(\omega) = -\frac{1}{\pi}\left[\text{Tr}\frac{(\tau_{0} + \tau_{3})}{2}\delta\hat{G}(\omega)\right].
 \end{aligned}
\end{equation}
Using Eq. \ref{eq:delRho}, we can compute anti-symmetrized impurity-induced LDOS $\delta\rho^{-}(\omega) = \delta\rho(\omega) - \delta\rho(-\omega)$ shown in Fig. \ref{fig:HAEMS}. Finally, we note that distinction between different gap symmetries on the basis of  $\delta\rho^{-}(\omega)$ occurs in $\omega$-space, therefore, it is sufficient to work at the level of lattice LDOS; continuum LDOS should yield the same conclusions as the lattice LDOS, as demonstrated in Ref. \cite{Boeker2020}.
%\subsection*{Appendix A}
%\subsection*{Appendix B}

%\begin{figure*}
%\begin{center}
%\includegraphics[width=2\columnwidth]{fig_cLDOS_DiS.pdf}
%\caption{Distinct spectral line shapes displayed by continuum LDOS in $d+is$-wave state. Orbital-resolved continuum LDOS in the proposed $d$-wave state (normalized with respect to the Fermi surface LDOS) at a number of intra-unit cell points in x-y plane located at height $z = 0.1c$ above NiO-plane (a)-(d) and R-plane (e)-(h). Inset shows position of the intra-unit cell point (red dot). Dashed vertical lines indicate two sets of coherence peaks at $\omega = \pm3.9$meV and $\omega = \pm2.35$meV.}
%\label{fig:cLDOS_d_is}
%\end{center}
%\end{figure*}

\bibliography{references}{}

\end{document}